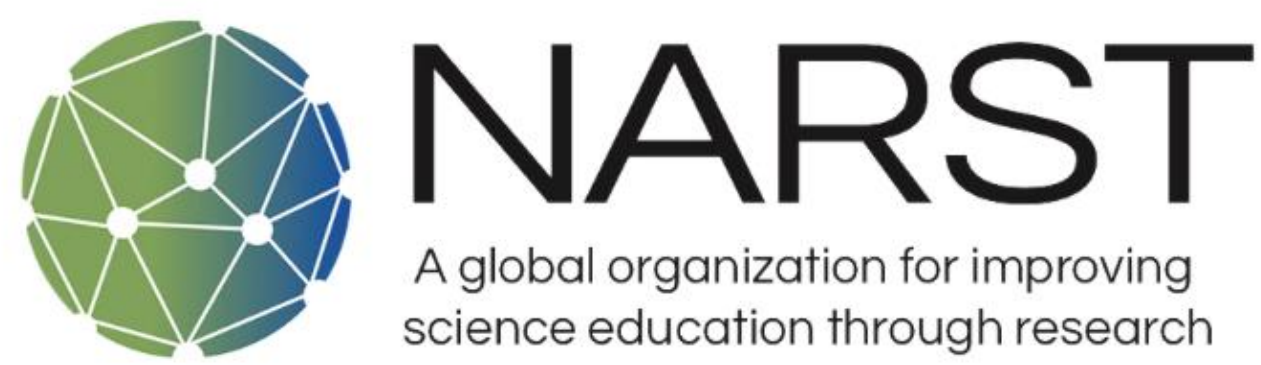

**Associations Between Event-Based (Mis)Recognition by STEM Authorities with STEM Identity and STEM Career Aspirations**

**Author(s):** Amdad Ahmed Awsaf, Florida International University; Remy Dou, Florida International University; Gerhard Sonnert, Harvard University; Philip Sadler, Harvard University

**Abstract**

Several programs for marginalized young people have been intentionally designed to increase the diversity in STEM careers; however, the participation of racially and ethnically minoritized individuals continued to be underrepresented in STEM disciplines. Using discipline-based STEM identity theories and drawing data from an NSF-funded survey of 1,134 participants, these regression analyses examine the impact of prior educational experiences on STEM identity (i.e., seeing oneself as a STEM person) and career aspirations among individuals attending Minority Serving Institutions (MSIs). The study's findings indicate that although positive reinforcement positively correlates with the STEM identity construct, the efforts are not always supportive enough to predict STEM career aspirations for minoritized individuals. It underscores the importance of explicitly designing appropriate interventions to support STEM identity formation and STEM career pursuit.

## Subject/Problem

Numerous research has consistently delineated how individuals' early life experiences, both in and out of school contexts, significantly impact their career choices (e.g., Maltese et al., 2014; Gossen & Ivey, 2023). Chen and colleagues (2024) asserted that students' self-perception as scientists positively correlates with their intention to pursue a STEM career[1]. Carrying out work at a Hispanic Serving Institution (HSI), Dou and Cian (2021) reported that contextual STEM learning in formal and informal educational settings can shape an individual's academic performance, career choice, and persistence in STEM-related fields. Much of this research points to the importance of experiences that inform individuals' sense of recognition (or misrecognition) from others (e.g., family, friends, teachers; Simpson et al., 2020; Avraamidou, 2022). However, little is known about the general contexts and characteristics of experiences that significantly contribute to their career aspirations (e.g., Dou et al., 2019). Such work would allow for the development of interventions that recognize young people as capable of succeeding in STEM fields. In this study, our objective was to examine experiences of public recognition (or misrecognition) of STEM performances in a population of undergraduate students pursuing a variety of majors at an HSI and the extent to which those experiences relate to their STEM identity and career aspirations.

Gossen and Ivey (2023) claimed that perceived learning experiences are associated with STEM self-efficacy and career aspirations. Self-efficacy, that is, confidence in one's ability to successfully accomplish tasks, might result from individuals' personal experiences and/or encouragement from others (Bandura, 1977). In the context of STEM, self-efficacy can effectively influence persistence in the STEM disciplines (Rittmayer & Beier, 2008). MacPhee and colleagues (2013) argued that minoritized students are often more likely to exhibit low STEM self-efficacy compared to others, which could stem from a lack of recognition from others (Li & Singh, 2023). However, STEM identity development is well understood to be shaped by individuals' perceptions of being recognized as a STEM person (e.g., Rodriguez et al., 2019), though much of this work also suggests that individuals from marginalized groups tend to experience misrecognition or lack of recognition in STEM contexts (Avraamidou, 2022; Carlone & Johnson, 2007; Rodriguez et al., 2019). Although learning experiences might vary depending on the learning context, Sahin et al. (2017) advocated that measuring the impact of perceived learning experiences might be an insightful approach to understanding STEM career aspirations.

---

[1] Science, Technology, Engineering and Mathematics (STEM) professions include medical doctors, health professionals, life scientists, astronomers, earth/environmental scientists, physical scientists, computer scientists, engineers, mathematicians/statisticians, and science or math teachers.

## Theoretical Framework

In their framework of students' identification with physics, Hazari and colleagues (2010) addressed the concept of discipline-based identity, highlighting three key precursors (i.e., recognition, interest, and performance–competence) that contribute to the development of one's STEM identity. Among these three components, "recognition" is considered the most prominent (Dou & Cian, 2022). Although research indicates that students' performance–competence (e.g., STEM self-efficacy) in STEM significantly impacts their STEM identities more than their interest in STEM alone (Godwin et al., 2016), a strong STEM identity (i.e., seeing oneself as a STEM person) is positively associated with an individual's aspiration to pursue a STEM career in the future (Dou et al., 2019). Simpson and colleagues (2020) also claimed that STEM identity plays a key role in shaping one's choice and persistence in the STEM field.

Considering the intricate relationship between perceived experiences, STEM identity, and STEM career aspiration, we sought to answer the following question,

1. To what extent do reported experiences of recognition or misrecognition in STEM contexts relate to the STEM career aspirations of racially and ethnically underrepresented youth?
2. To what extent do experiences of STEM recognition or misrecognition shape racially and ethnically underrepresented youth's STEM identities?

## Research Design/Procedure

### Data sources

Drawing on data from an NSF-funded survey of racially and ethnically minoritized youth's out-of-school-time experiences and career choices, we examined the impact of various educational experiences on respondents' STEM identity and aspirations to choose a STEM career in college. The participants of the study ($N$ = 1,134) were selected using convenience sampling. With support from a coalition of Minority Serving Institutions (MSIs; e.g., Historically Black Colleges, HSIs, and Tribal Colleges and Universities), college English instructors were invited to distribute the survey to their students. Among the respondents, 67% were first-year students, 15% were second-year, and 6% were other students. In terms of gender, 50% self-identified as female, 38% as male, and 1.8% as non-binary. The remaining group of respondents either did not specify or self-describe their gender. Moreover, 48% of respondents self-identified as Black, 34% as Hispanic, 31% as white, 3% as Asian, and 2% as American Indian or Alaskan Native. Respondents were afforded the agency to identify with multiple races or ethnicities, including groups not included in our list of options, which mirrored the U.S. Census 2020 (e.g., Sarraju et al., 2023).

**Survey Items**

The survey included questions that invited students to report whether they were “often called upon to answer questions in their STEM classes”, “recommended by teachers for an advanced STEM class”, “invited by adults to participate in STEM competitions”, “accepted into STEM programs or school that required an application” and “received awards while participating in STEM activities”. Misrecognition experiences were measured in a similar manner using the following items: “at least one teacher did not recommend for an advanced STEM class that they believed they belong in”, “had a negative learning experience during an out-of-school STEM program”, and “had a negative learning experience with STEM at school”. This collection of survey items was underpinned by a review of literature (e.g., Dou et al., 2019; Roberts et al., 2018, Park et al., 2014) and reviewed by a panel of STEM education experts (Anonymized NSF Award No.).

We used a previously validated, single-item question to measure respondents’ identification with STEM fields (Dou & Cian, 2022). Students were instructed to rate their level of agreement on a 5-point Likert scale with the following statement: “I see myself as a STEM person”. STEM career aspiration was measured by assessing respondents' aspiration to various STEM professions. If respondents expressed interest in any STEM profession at the "end of high school" or "beginning of college", they were classified as STEM career aspirants; otherwise, as non-STEM aspirants[2]. Given that prior research has indicated a strong association between

[2] Non-STEM professions include other teachers, anthropologists/archaeologists, social scientists, humanities professionals, visual/performing artists, business persons, lawyers, politicians, athletes/coaches, military personnel, and other non-STEM related careers.

STEM identity and student gender, as well as their reported racial identities, we included additional items inviting students to self-identify their gender and racial/ethnic identity.

## Analysis Procedures

We conducted stepwise multiple linear and logistic regression analyses to investigate the association between students' reported experiences and their association with STEM identity and STEM career aspirations. We used logistic regression models to test the relationship between STEM career aspiration and responses to the recognition and misrecognition items while controlling for respondent gender and racial identities. These latter variables were not found to be significantly associated with STEM career aspirations. Similarly, we tested multiple linear regression models to determine the association between respondents' STEM identity and their prior experiences. The results of our model testing suggested the potential for multicollinearity between *"answering questions*" and "*receiving awards*"; therefore, we conducted a Principal Components Analysis (PCA) (e.g., Kelechi, 2012). The biplot generated from PCA (Figure 1) illustrates those two principal components, RC1 and RC2, explaining 27% and 16% of the total variance, respectively. The first principal component (RC1) predominantly captured the variance associated with reported STEM recognition experiences (*i.e., answering questions, teacher's recommendation, adult's invite, being accepted in STEM programs, and receiving awards*). In contrast, the second component, RC2, was associated with reported misrecognition experiences (*i.e., no teacher recommendation, negative out-of-school experience, negative school experience*). In the subsequent stages, we conducted regression analyses (e.g., Schommer-Aikins et al., 2005) to determine the influence of reporting an experience related to at least one of the recognition items and reporting an experience related to at least one misrecognition item.

## Findings

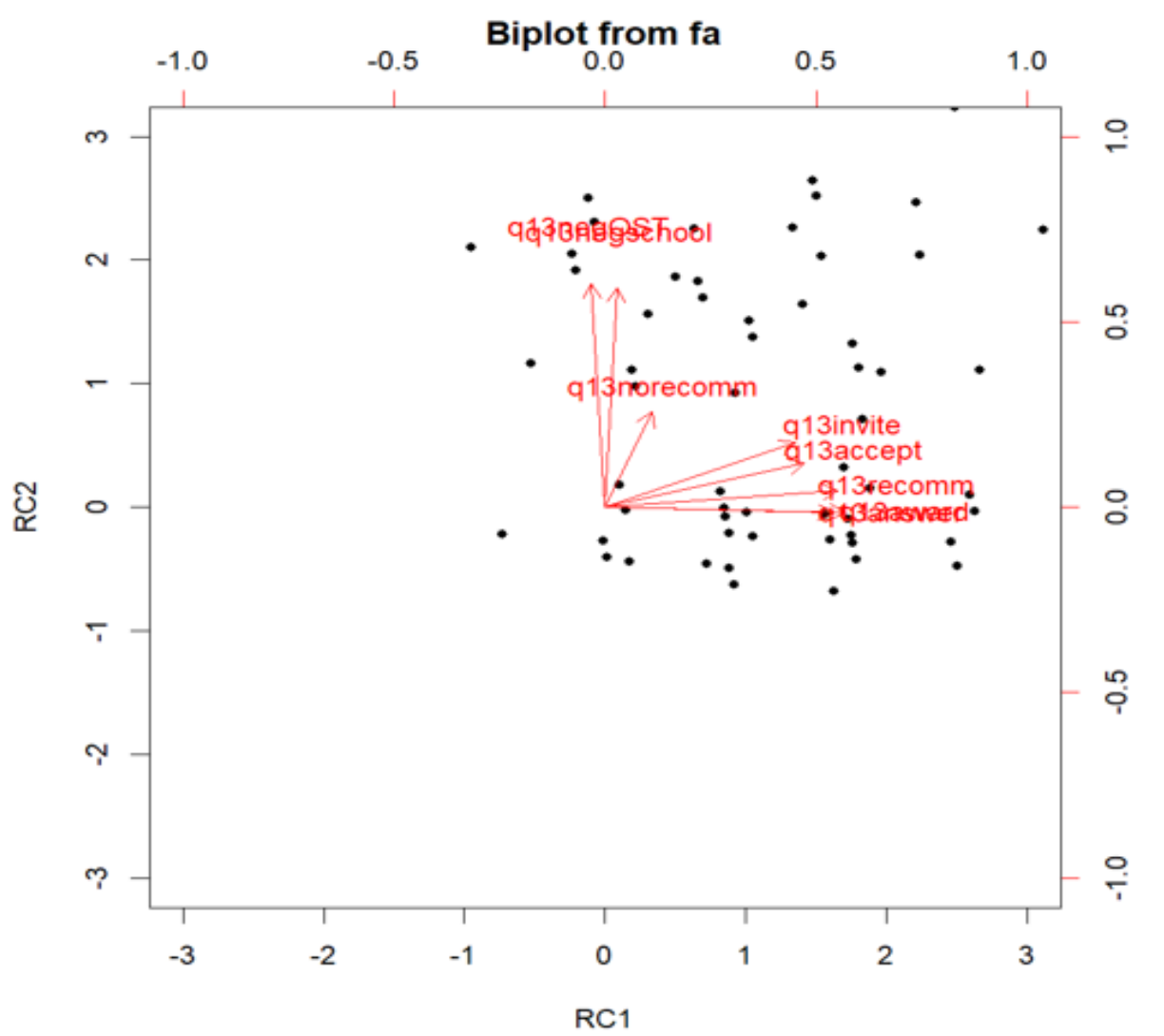

*Figure 1: Biplot from Principal Components Analysis (PCA)*

Our analyses revealed that having at least one of the following recognition experiences, that is, getting called upon to answer questions in their STEM classes, being recommended by a teacher for an advanced STEM class, and receiving awards while participating in STEM activities, was a statistically significant predictor of STEM-related career aspirations in college (β = 0.47, SE = 0.15, *p* = .002). Specifically, individuals who reported having at least one of those positive experiences prior to college exhibited 1.59 higher odds of aspiring to a STEM career. Interestingly, having the experience of being invited by adults to participate in a STEM competition and getting accepted into a STEM program or school that required an application was not statistically significant in predicting STEM career aspirations. Similarly, neither demographic variables (i.e., gender and race) nor misrecognition experiences exhibited statistically significant associations with our outcome variable.

Furthermore, we tested multiple linear regression models to determine the relationship between reported experiences and students' self-identification as STEM persons. We graphed our continuous variables using scatter plots and histograms to identify potential outliers and calculated skewness and kurtosis values where appropriate to identify potential issues related to heteroscedasticity. The outcomes of these analyses did not suggest meaningful violations of the assumptions of linear regressions. Observations with missing values were removed using listwise deletion (354 observations were removed due to missing values). Given the high level of missingness, we present our findings as preliminary and intend to use multiple imputations (Rubin, 1988) to address missingness in future iterations of these analyses. Our final regression model was statistically significant ($R^2 = 0.1342$, $F[12, 780] = 10.08$, $p < 0.001$), finding that being recommended by a teacher for an advanced STEM class ($\beta = 0.7029$, $p < 0.001$), receiving awards while participating in STEM activities($\beta = 0.6701$, $p < 0.001$), getting called upon to answer questions in STEM classes ($\beta = 0.4318$, $p = 0.01$), and getting accepted into a STEM program or school that required an application is ($\beta = 0.6881$, $p = 0.003$) were statistically significant predictors of student's identification with STEM at the time of the survey. On the contrary, being invited by adults to participate in a STEM competition or having a misrecognition experience were not significant predictors of students' identification with STEM.

## Contributions & Implications

STEM identity plays a key role in underrepresented youth's success in STEM (Carlone & Johnson, 2007; Dou & Cian, 2021) and "recognition" is a positive approach to acknowledging success by others, which can positively shape one's STEM identity (e.g., Mann, 1988). Although prior studies (e.g., Dou et al., 2023; McDavid et al., 2020) have delineated that various positive reinforcement and recognition experiences have positive associations with individuals' STEM interests and competence, our analysis explicitly indicates that receiving awards while participating in STEM activities, being recommended by a teacher to an advance STEM course, and getting called upon to answer questions in STEM classes have significant associations with students' STEM career aspirations and identity in college. Though preliminary, these results have implications for both K-12 and out-of-school learning settings for guiding stakeholders in developing efficacious interventions aiming to increase participants' STEM career aspirations and identification with STEM fields. Given our findings, interventions that focus on fostering one or more of these three positive recognition experiences would support students who identify as female and/or with historically underrepresented racial and ethnic groups (e.g., Habig et al., 2018).

Interestingly, our findings also suggest that misrecognition experiences (i.e., *no teacher recommendation, negative out-of-school experience, negative school experience*) may not be associated with individuals' STEM career aspirations or identity, which contrasts with previous research findings (e.g., Chachashvili-Bolotin et al., 2016; Simpson & Maltese, 2016), where the researchers argue that negative learning experience might function as an impediment for one's aspiration to pursue a STEM career. Our findings underscore the significance of further exploration of other factors that might potentially impact STEM identity development (e.g., Patrick et al., 2018) and career aspirations. In the context of the NARST 2025 Conference Theme, it is perhaps more important to note that teachers play a critical role by acknowledging and recognizing student's success in STEM-related tasks, which is positively associated with STEM identity and career aspiration. Therefore, the NARST members who are concerned with advancing equitable STEM education may adopt collaborative efforts to research with teachers in determining effective approaches and co-develop interventions that acknowledge various STEM recognition, particularly for those who are traditionally underrepresented in STEM (e.g., women and ethnic minorities). It may also benefit NARST members to extend research on reframing and reforming existing recognition approaches in formal and informal STEM learning contexts to ensure the development of a robust STEM identity and STEM career aspiration for minority populations.